# Low Latency Real-Time Seizure Detection Using Transfer Deep Learning

*V. Khalkhali[1], N. Shawki[1], V. Shah[1], M. Golmohammadi[2], I. Obeid[1], and J. Picone[1]*

1. Neural Engineering Data Consortium, Temple University, Philadelphia, Pennsylvania, USA
2. Internet Brands, El Segundo, California, USA
{vahid, nabila.shawki, vinitshah, meysam.gol.m, obeid, picone}@temple.edu

**Abstract— Scalp electroencephalogram (EEG) signals inherently have a low signal-to-noise ratio due to the way the signal is electrically transduced. Temporal and spatial information must be exploited to achieve accurate detection of seizure events. Most popular approaches to seizure detection using deep learning do not jointly model this information or require multiple passes over the signal, which makes the systems inherently non-causal. In this paper, we exploit both simultaneously by converting the multichannel signal to a grayscale image and using transfer learning to achieve high performance. The proposed system is trained end-to-end with only very simple pre- and post-processing operations which are computationally lightweight and have low latency, making them conducive to clinical applications that require real-time processing. We have achieved a performance of 42.05% sensitivity with 5.78 false alarm per 24 hours on the development dataset of v1.5.2 of the Temple University Hospital Seizure Detection Corpus. On a single core CPU operating at 1.7 GHz, the system runs faster than real-time (0.58 xRT), uses 16 Gbytes of memory, and has a latency of 300 msec.**

## I. Introduction

The electroencephalogram (EEG) is still the primary tool used in hospital clinical and critical care settings to diagnose brain-related illnesses. Epilepsy is the fourth most common neurological problem – only migraine, stroke, and Alzheimer's disease occur more frequently [1]. In 2015, 1.2% of the US population were diagnosed with epilepsy (3 million adults and 470,000 children) [2]. Obviously, this large population of patients cannot be monitored by physicians continuously. This is particularly true for patients who undergo long-term monitoring during in-patient care. Long-term monitoring can last more than 24 hours creating an enormous amount of data that must be manually reviewed. Ambulatory EEGs, which we do not address in this study, generate significantly more data and pose even greater challenges. Hence, there is a great need for accurate automated seizure detection. Seizure prediction, which involves prediction of a seizure event before it happens, is even more important and is an emerging field.

There have been many attempts to design systems that can detect seizures from noninvasive EEG signals [3][4]. However, these high performing systems are often non-causal and/or non-real time (NRT) and have significant amounts of latency because they perform multiple passes over the signal. These systems typically preprocess the signal, perform feature extraction, and then use several deep learning approaches to process these features. Postprocessing of hypotheses always seems to play a big role in achieving high performance but introduces significant amounts of latency. Sharmila et al. [5] have done a comprehensive review and concluded that feature extraction using a discrete wavelet transform (DWT) is the dominant feature extraction approach in this field.

Golmohammadi [6] and Shah [7] proposed hybrid architectures which used linear frequency cepstral coefficients (LFCCs) as features followed by convolutional (CNNs) and long short-term memory networks (LSTM). LFCCs are a filter bank-based representation that has been successful in many other signal processing applications prior to the introduction of deep learning systems. Their methods classify the EEG signal into three seizure patterns (e.g., spikes, generalized periodic epileptiform discharges) and three background categories (e.g., eye movement, artifacts, and background). Their sensitivity and specificity were approximately 90% and 95%, respectively.

Craley et al. [8] exploited a hybrid Probabilistic Graphical Model CNN (PGM-CNN) for seizure tracking. They used an engineered feature called a Coupled Hidden Markov Model (CHMM) that is an extension of conventional Hidden Markov Models where the current state is not only dependent on the states of its own chain but also depends on the neighboring chain at the previous time-step. The classification results are evaluated on a Johns Hopkins University Hospital dataset (JHH) that contains 90 seizures from 15 patients [9] and the Children's Hospital of Boston (CHB) dataset that has 185 recordings from 24 pediatric patients [10]. Their proposed method achieved a true positive rate and false positive rate of 45% and 1% on JHH, and 61% and 1.3% on CHB, respectively.

Emami et al. [11] proposed an end-to-end seizure detection system without the need for feature extraction. First, they filter the raw EEG signals with a bandpass and a notch filter. Then they converted this multichannel time series to an image, replicating what neurologists used to manually interpret an EEG. Next, a CNN was employed to detect seizure events. They achieved a true positive rate of 74% with a false alarm rate of 0.2 per hour on data from eight subjects collected at the NTT Medical Center Tokyo and 16 subjects collected at the University of Tokyo Hospital.

Gomes et al. [12] proposed another end-to-end seizure detection system based on processing EEGs as images using CNNs. They exploited data augmentation by using small shifts of overlapping windows for training. They evaluated their approach on two datasets: CHB and the European Epilepsy Database (EPILEPSIAE) [13]. On CHB, they achieved an accuracy of 99.3%, a specificity of 99.6%, and a false alarm rate of 0.5 per hour for 92% of the patients. On EPILEPSIAE, their accuracy and specificity were 98.0% and 98.3%, respectively, with 1.0 per hour false alarm rate for 80% of the patients.

In this paper, we propose the application of transfer learning to the seizure detection problem, as shown in Figure 1. We will focus on seizure detection on the Temple University Hospital Seizure Database (TUSZ) [14]. Since high quality annotated data is in short supply, transfer learning can provide a more efficient and effective way to train a neural network.

## II. Application of Transfer Learning

Manual interpretation of an EEG signal requires detection of very subtle pattens. Accurate classification of these patterns requires a deep learning system with many parameters, which in turn, requires large amounts of manually annotated training data. When the number of model parameters greatly exceeds the number of patterns available for training, overfitting becomes a huge concern. Dropout, batch normalization and other regularization methods attempt to mitigate this, but a better approach is to add relevant data from other sources.

Neurologists are capable of manually interpreting EEGs with accuracies that can exceed machine performance [6][15]. Hence, we are confident that there is adequate information content in the signal, particularly visualizations of the waveforms, to classify seizures. Image processing approaches are emerging, such as those disclosed in ImageNet competitions [16], that attempt to emulate the way humans interpret visual data in an application independent manner. In this way, we can leverage vast amount of image training data available from other applications, such as object recognition and autonomous vehicle navigation. In this paper, we attempt to leverage these pretrained models from the ImageNet competition and adapt them to seizure detection.

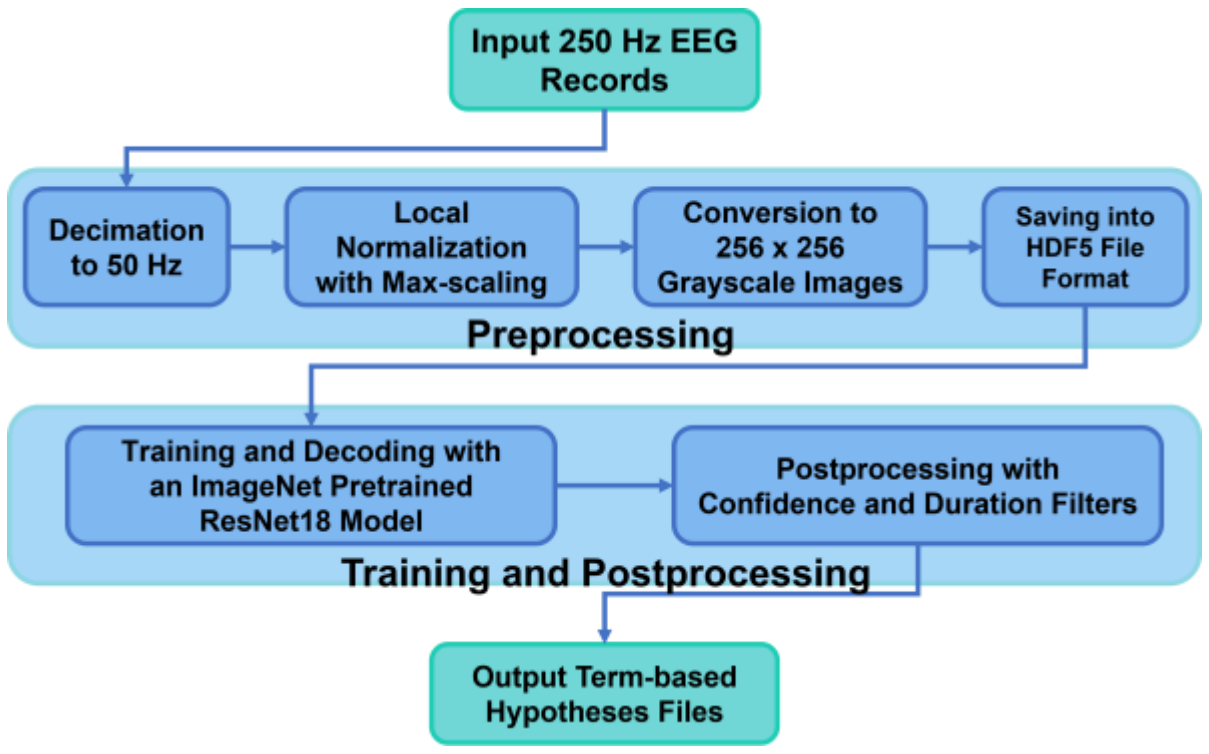

Figure 1. An image processing approach to seizure detection

One of the well-known and efficient networks that has achieved good performance is ResNet18 [17]. It is a type of deep residual network that overcame the limitations of training very deep networks by introducing identity shortcut connections between the layers. The ResNet18 model consists of four module blocks as shown in Figure 2. Key parameters for each layer are given in Figure 3. The input block contains a two-dimensional CNN followed by batch normalization and max pooling. The output block consists of an average pooling operation that improves generalization and a linear layer that performs classification. The parameter (64,64) in Figure 3 refers to the dimensions of the input and output planes (channels), respectively.

The four hidden layers are similar in design. The first layer consists of two blocks referred to as Basic blocks. Each of these Basic blocks contains a CNN, a batch normalization block, and a second CNN. The remaining three layers contain a Basic block, followed by a downsample block, followed by a second Basic block. The downsample block adapts the input and output sizes of the Basic blocks so that they can be concatenated. The network is loaded with pretrained parameters from ImageNet. During training, we optimize parameters using a cross-entropy loss function with predefined prior probabilities (weights).

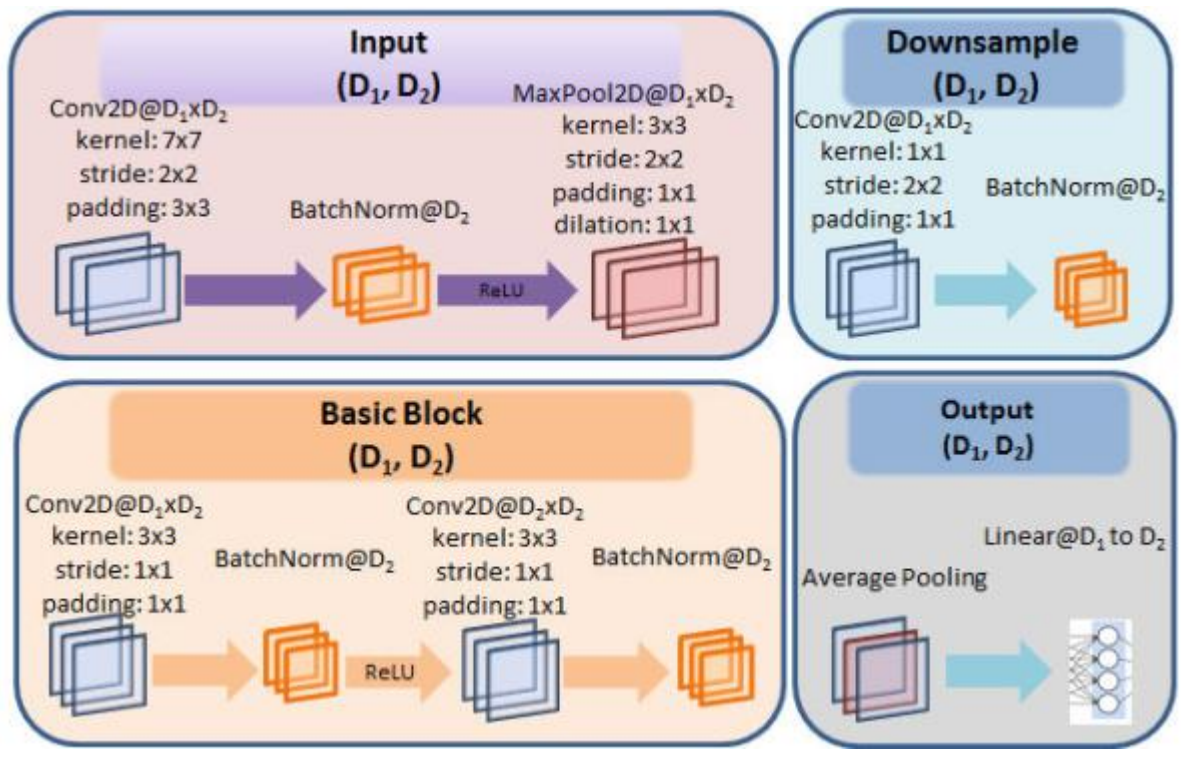

Figure 2. Construction blocks of the ResNet-18 architecture

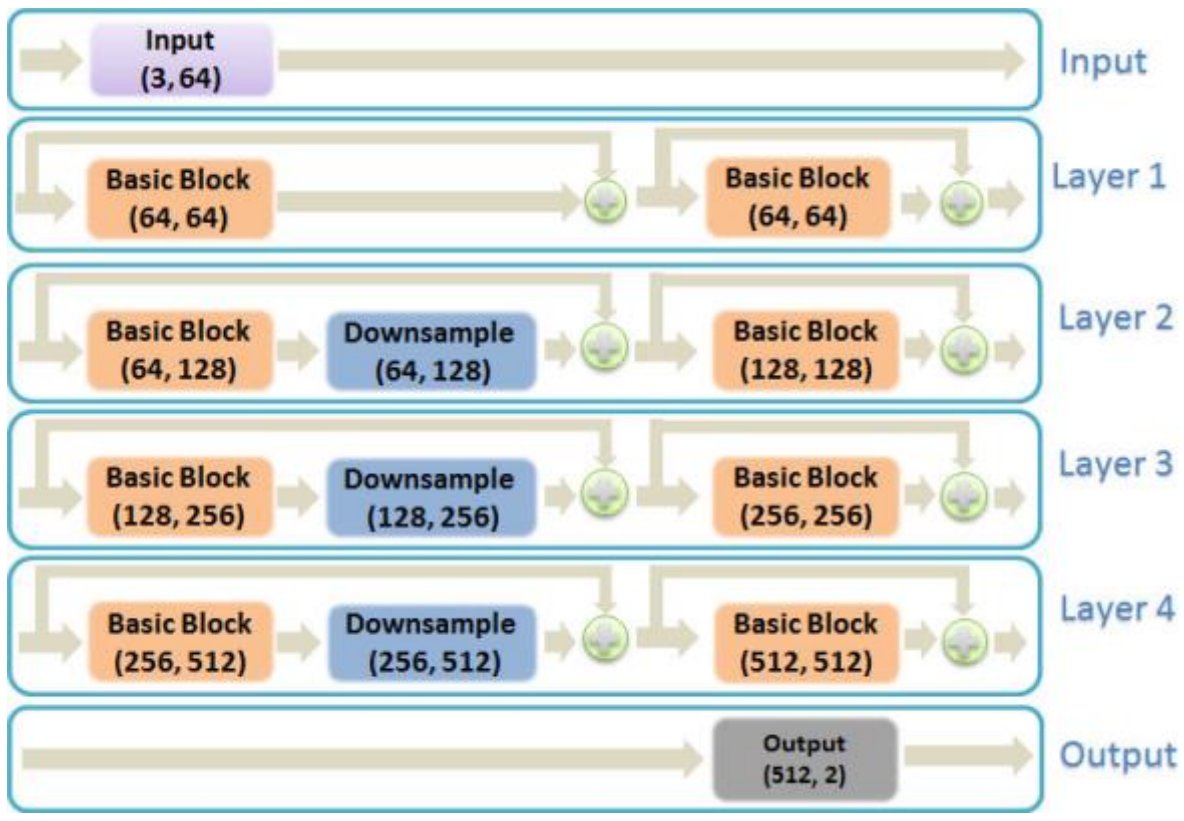

Figure 3. An overview of the ResNet-18 architecture

Seizures occur about 7% of the time in the TUSZ Corpus. Since the distribution of classes is highly unbalanced, we must adjust the weights of the loss function. Hence, to better balance the number of seizure and background patterns, we randomly selected one-fifth of the background patterns. It was observed that using all the background samples did not improve the results. In practice, we have used about 30,000 samples for seizure events and 100,000 samples for background events in an HDF5 database format [18]. Since, the samples in HDF5 files are arranged sequentially, random sampling (without replacement) removes the need for a windowing process with predefined overlaps.

To further adjust for the unbalanced number of samples in the two classes, we include prior probabilities in the loss function computation:

$$\omega_b = \frac{N_{seiz}}{N}, \quad \omega_s = \frac{N_{bckg}}{N}, \tag{1}$$

where $N$, $N_{seiz}$, and $N_{bckg}$ are the total number of samples, the number of seizure samples, and the number of background samples, respectively. Variables $\omega_b$ and $\omega_s$ are the background and seizure weights, respectively.

The loss function is defined as:

$$loss_{weighted} = \omega_b * loss(x_i, b) + \omega_s * loss(x_i, s), \tag{2}$$

where $x_i$, $b$, $s$ are the $i^{th}$ input sample, the background class index, the seizure class index, respectively. The function $loss(x, y)$ is the balanced loss function, typically implemented as cross-entropy or mean square error. The weighted loss function defined in (2) alleviates the unbalanced number of samples in the two classes and avoids biased learning.

Stochastic gradient descent (SGD) with an adaptive step size is used with 25 epochs and a batch size of 8. During the training process, random resized crop and random horizontal flip transformations, similar to what is used in the training process for this system in ImageNet [16], were exploited to force the network to learn spatiotemporal information.

## III. Experimentation

Our focus in this study was TUSZ v1.5.2, which was used in the Neureka™ 2020 Epilepsy Challenge [3]. Some relevant statistics of the corpus are shown in Table 1. The EEG signals in TUSZ are preprocessed using montages to suppress background noise and accentuate spikes [19].

Table 1. TUSZ v1.5.2 statistics

| Description | Train | Dev |
|---|---|---|
| Patients | 592 | 50 |
| Sessions | 1247 | 342 |
| Files | 5521 | 1656 |
| Files with Seizure | 840 | 246 |
| Patients with Seizure | 199 | 39 |
| Seizure Events | 2332 | 599 |
| Total Dur (sec.) | 2,910,639 | 1,598,493 |
| Seizure Duration Ratio | 6.34% | 8.96% |

These montages produce 21 differential signals extracted from 19 raw channels. The channels are decimated to 50 Hz since there is little useful information above 25 Hz and this reduces the processing time considerably.

Scaling or normalization of the resulting signals is a very important next step. Channel dependent or global normalization does not work well for this data. Moving average filters, which are also popular, can significantly diminish spike behavior. Outlier removal is another popular method for artifact removal, but in practice this approach also tends to remove many seizure events. We chose to implement a local scaling approach, which we refer to as max local scaling, in which all samples in a window are scaled between $[-1, +1]$:

$$A_{max}[n] = \max(|A[i]|)\,;\; n - {}^{N}\!/_{2} \le i \le n + {}^{N}\!/_{2}\,, \tag{3}$$

$$\hat{A}[n] = A[n]/A_{max}[n]. \tag{4}$$

where $A$ is amplitude, $n$ is the current sample and $N$ is the number of samples in a window centered around the current sample. $\hat{A}[n]$ is the locally scaled sample.

In our work, we found that implementing local scaling with a 6-second center-aligned window gives good results. Figure 4 shows two windows with spikes that have been scaled. The orange waveform corresponds to the original signal and blue waveform represents the rescaled signal.

Note that this type of scaling removes the absolute energy of the signal as a feature. In some types of seizures, the evolution of energy over time is important. We could add energy back as an additional feature, but in the current work we treat each window as an independent sample. We use a postprocessor to consider the sequences of events and make a final decision based on the evolution over time of these events.

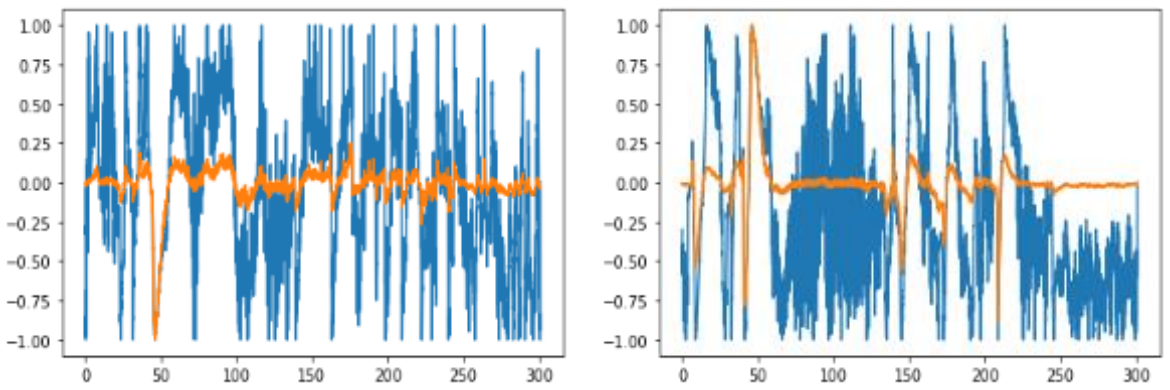

Figure 4. Window normalization with max local scaling

The next step for data preparation is converting all samples into grayscaled images. Since in the previous step EEG signals have been scaled such that the amplitudes are limited to the range $[-1, +1]$, conversion to a grayscaled image where pixels fall in the range $[0, 255]$ is straightforward. Through experimentation, we found using $256$ samples for every window for every channel is efficient. But, since it is necessary to use a square shaped image as input to the ImageNet system, and the height of windows is equal to the number of montages, then rescaling these images is necessary. Therefore, the windows are resized from $20 \times 256$ to $256 \times 256$ with cubic interpolation. Several examples of the resulting images are shown in Figure 5. The confusion matrices for ResNet18 are shown in Table 2 for the training data (closed-loop training) and the development data (open-loop training) Table 3.

## IV. Postprocessing

As we have stated, seizure detection is a challenging problem. The raw classification performance of ResNet18 is poor and needs additional postprocessing. Our postprocessor is based on three parameters:

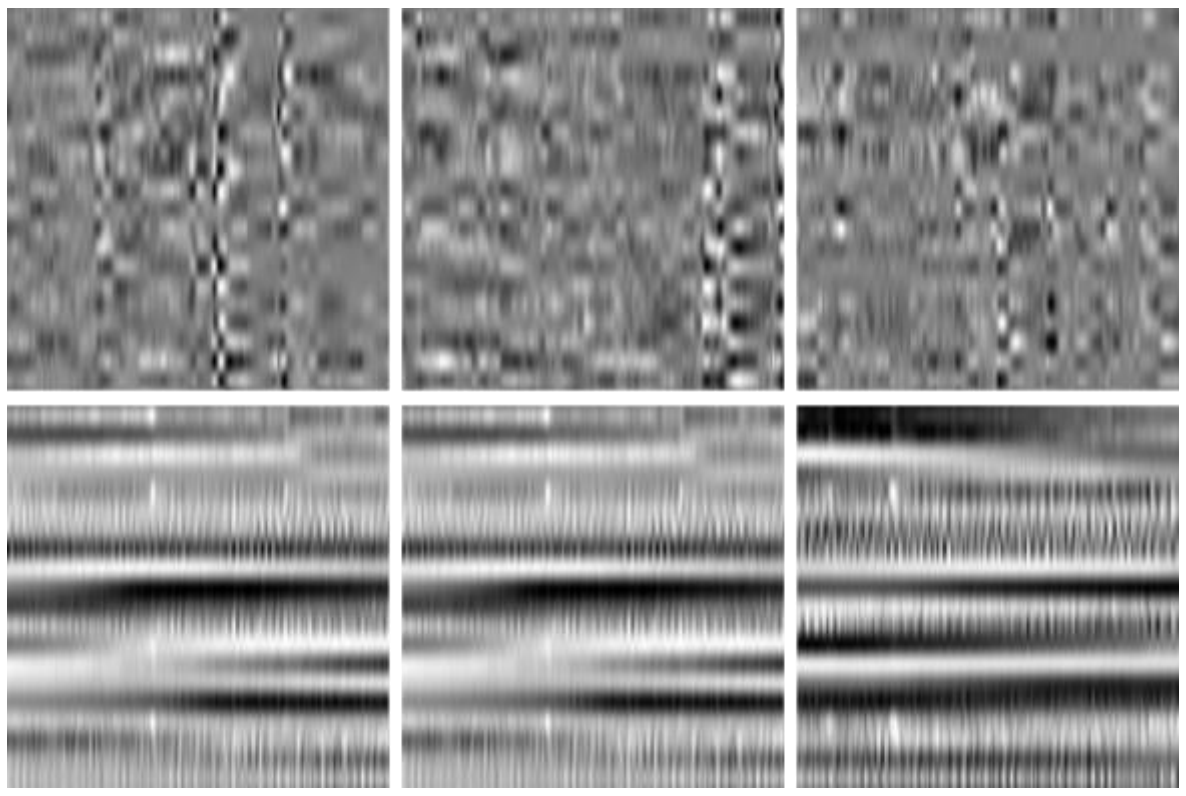

Figure 5. Grayscale images of background windows (first row) and seizure windows (second row)

Table 2. The confusion matrix for the training dataset

| Training | | Detected | |
|---|---|---|---|
| | | Background | Seizure |
| **Actual** | Background | 84.67% | 17.30% |
| | Seizure | 20.35% | 79.65% |

Table 3. The confusion matrix for the development dataset

| Development | | Detected | |
|---|---|---|---|
| | | Background | Seizure |
| **Actual** | Background | 82.70% | 17.30% |
| | Seizure | 39.53% | 60.47% |

- Seizure confidence threshold, $S_{th}$: events with probabilities less than this threshold are classified as background while events with higher probabilities are classified as seizures;
- Minimum acceptable background duration, $BD_{min}$: all background events with a duration less than this value are converted to seizure events;
- Minimum acceptable seizure duration, $SD_{min}$: all seizure events with a duration less than this value are converted to background.

The first parameter reduces noisy decisions by increasing the seizure confidence threshold. The second and third parameters act similar to dilation and erosion in morphological image processing [20]. Once the threshold $S_{th}$ is applied, then all background windows between two seizure events with a duration less than $BD_{min}$ will be classified as seizures. Similarly, all seizure events with a duration less than $SD_{min}$ will be classified as background.

Postprocessing has a significant impact on both the misclassification errors and the false alarm rate. Figure 6 shows the sensitivity as a function of the detection latency. Figure 7 shows the false alarm rate as a function of the detection latency. Detection delay is defined as the summation of minimum acceptable duration of background and seizure events:

$$Detection\ Delay = BD_{min} + SD_{min}. \quad (5)$$

The false alarm rate is significantly reduced by increasing the detection delay. There is also a moderate reduction in sensitivity.

## V. Evaluation Results

Several metrics for evaluating the efficiency of EEG seizure detection systems are discussed extensively in Shah et al. [4]. In Figure 8 we compare the performance of the proposed system to two other previously published systems: a hybrid CNN/LSTM system (cnn_lstm) developed by Golmohammadi [6] and a multiphase system (mphase) developed by Shah [7]. This comparison was performed on the development data set (dev) using the OVLP scoring metric [21]. Note that we focus on performance for the range [0,0.1] where false alarms are very low. This is the operating region of most interest for this application. In Figure 9, a similar analysis is shown for the blind evaluation set (eval). It is important to note that no tuning was performed based on the evaluation set results.

The seizure sensitivity and false alarm rate of the new system, which are 42.05% and 5.78/24h, respectively, are marginally better than our previous best results (40.12% and 6.62/24h for the multiphase system). However, it is important to note that the resnet system is much simpler, has much less latency, and runs faster than real-time on relatively modest computing resources. This makes it ideal for critical care applications.

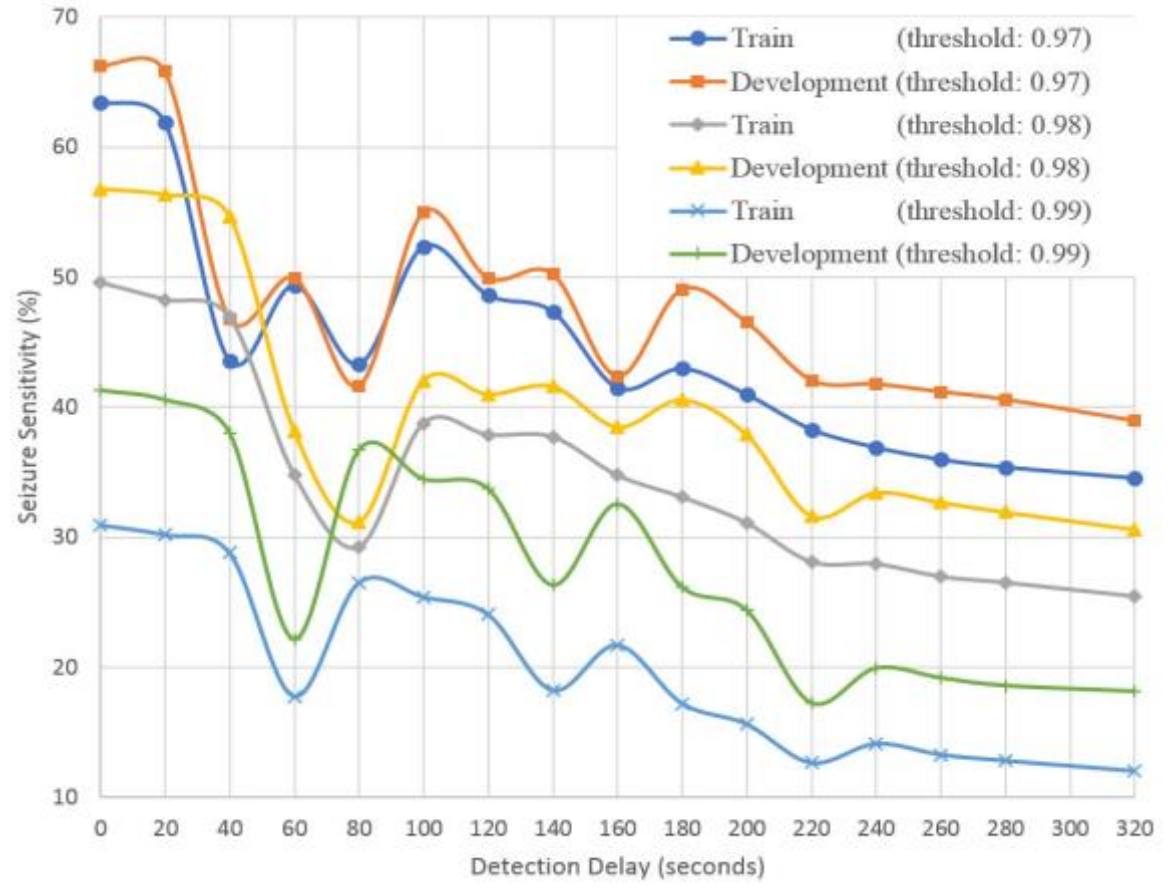

Figure 6. Sensitivity as a function of delay

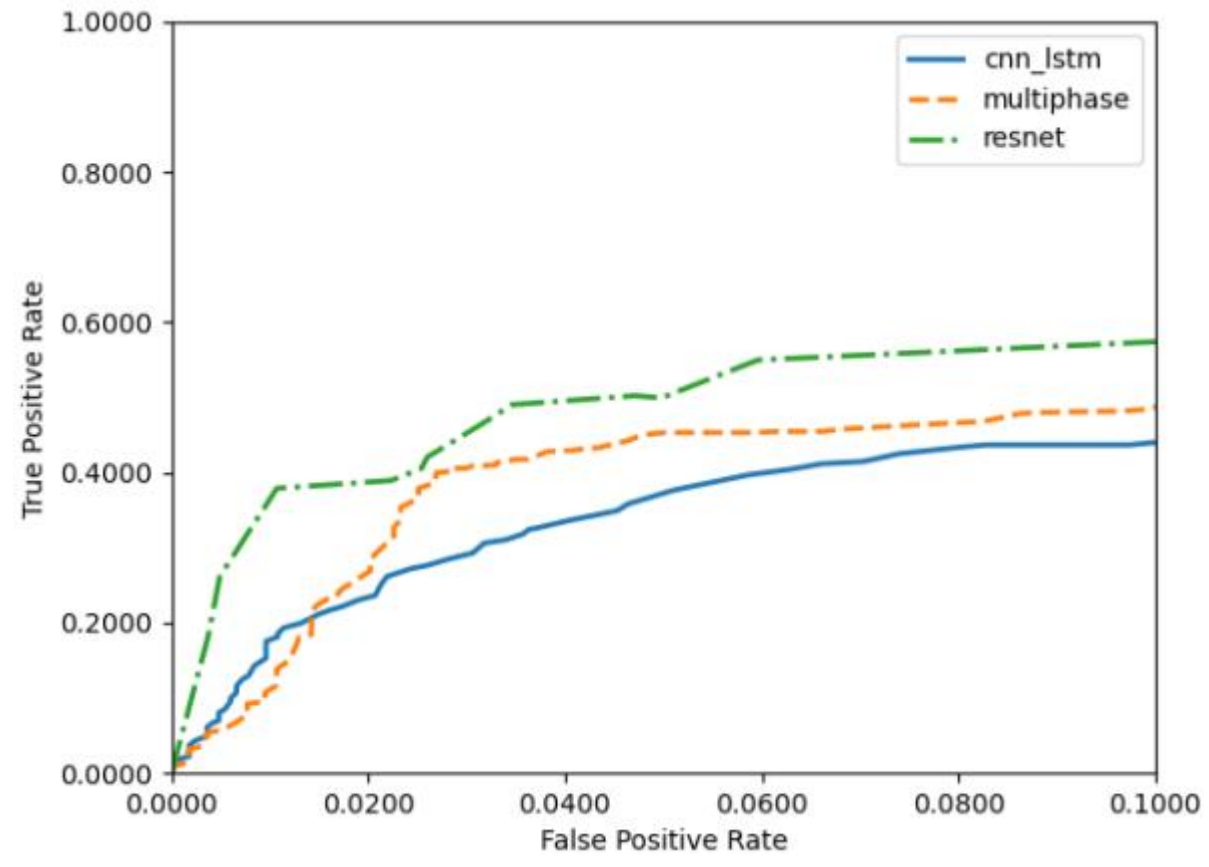

Figure 8. An ROC comparison on the development dataset

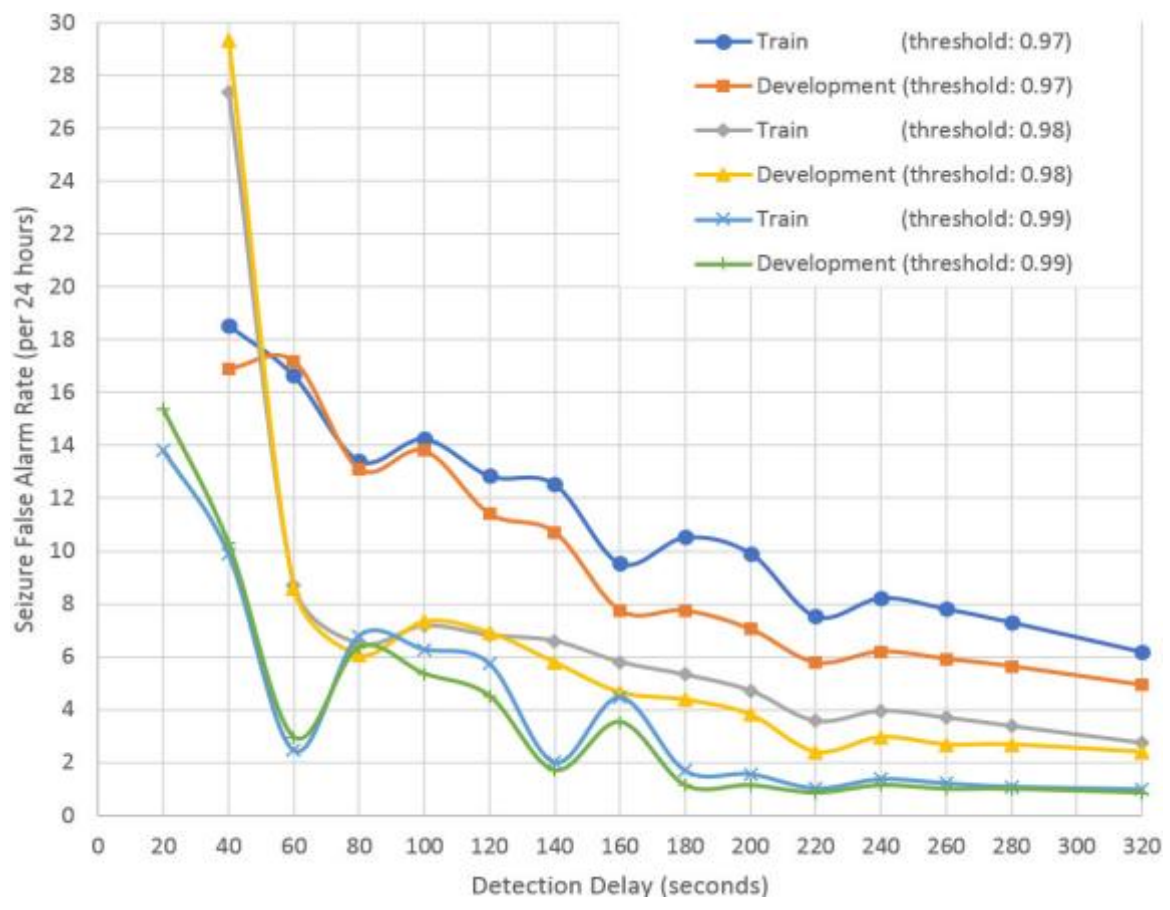

Figure 7. False alarm rate as a function of delay

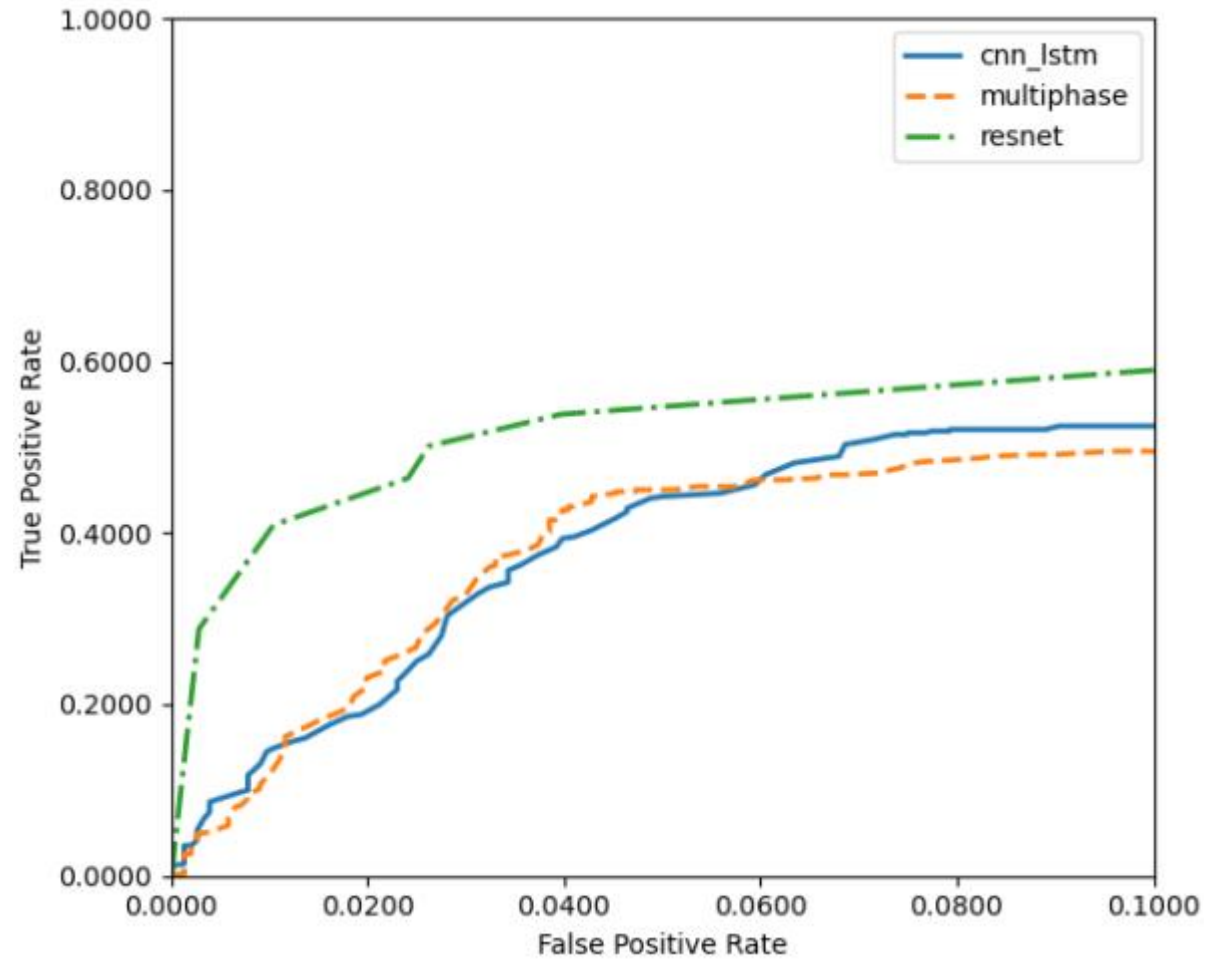

Figure 9. An ROC comparison on the evaluation dataset

In Table 4 and Table 5, we compare the results of the proposed system to several leading systems including two of the top systems that participated in the Neureka™ 2020 Epilepsy Challenge [3]. These tables include the raw scores for sensitivity, specificity, and the false alarm rate for four scoring metrics. The last metric, time-aligned event scoring (TAES), also shows the final weighted score (WGT) using the weighting function adopted for the competition.

From these results, we see that the performance of the resnet system still lags the two best performing systems from the competition, sia and pnc98. However, the resnet system was designed to be real-time and low latency, while the best performing systems in the competition were non-real time with infinite latency and focused more on minimizing the weighted error metric. The competition's weighted error metric placed a great emphasis on minimizing the false alarm rate, since this is a crucial parameter in critical care applications, so it was prudent for these system developers to sacrifice sensitivity for the competition.

Also, the competition applied a penalty based on the number of channels used, encouraging participants to use as few channels as possible. We have not explored this dimension with the resnet system yet. Using a reduced number of channels is more relevant to consumer-grade applications. In critical care applications, all available channels are collected and processed, so aside from the computational considerations, there is no benefit to reducing the number of channels processed.

## VI. Conclusions

Designing high sensitivity, low false alarm, and low delay seizure detection systems from noninvasive electroencephalogram scalp measurements is challenging. We have shown that with some trade-offs between three critical parameters – sensitivity, false alarm rate, and detection delay – real-time performance can be achieved without sacrificing performance.

Table 4. A comparison to leading systems in the Neureka™ 2020 Epilepsy Challenge for the development data

| Metric | | cnnlstm | mphase | resnet | sia | pnc98 |
|---|---|---|---|---|---|---|
| DPAL | Sens | 45.17 | 36.70 | 20.36 | 23.45 | 6.98 |
| | Spec | 88.63 | 96.25 | 96.72 | 99.47 | 98.33 |
| | FPs | 23.25 | 6.76 | 5.50 | 0.97 | 2.54 |
| EPCH | Sens | 37.67 | 36.34 | 49.83 | 12.84 | 1.56 |
| | Spec | 96.56 | 97.16 | 92.64 | 99.97 | 99.99 |
| | FPs | 2686.97 | 2221.29 | 5753.08 | 25.85 | 8.45 |
| OVLP | Sens | 43.69 | 40.12 | 42.06 | 23.26 | 6.39 |
| | Spec | 91.71 | 97.19 | 97.40 | 99.74 | 99.65 |
| | FPs | 20.85 | 6.62 | 5.78 | 0.64 | 0.85 |
| TAES | Sens | 35.83 | 32.27 | 15.34 | 11.38 | 2.04 |
| | Spec | 83.91 | 90.18 | 88.81 | 99.46 | 99.42 |
| | FPs | 32.55 | 18.07 | 19.42 | 0.99 | 0.87 |
| | WGT | -53.05 | -21.21 | -40.71 | 2.59 | 0.83 |

Table 5. A comparison to leading systems in the Neureka™ 2020 Epilepsy Challenge for the evaluation data

| Metric | | cnnlstm | mphase | resnet | sia | pnc98 |
|---|---|---|---|---|---|---|
| DPAL | Sens | 54.79 | 45.01 | 14.29 | 24.07 | 8.61 |
| | Spec | 90.41 | 94.47 | 98.03 | 99.31 | 99.44 |
| | FPs | 21.79 | 11.61 | 3.50 | 1.27 | 0.95 |
| EPCH | Sens | 38.15 | 52.04 | 42.82 | 1.27 | 5.09 |
| | Spec | 98.40 | 98.27 | 95.88 | 99.95 | 100.00 |
| | FPs | 1282.04 | 1391.52 | 3314.04 | 43.23 | 2.07 |
| OVLP | Sens | 51.47 | 44.62 | 37.18 | 23.88 | 8.41 |
| | Spec | 92.63 | 95.48 | 98.33 | 99.61 | 99.93 |
| | FPs | 19.25 | 11.45 | 3.82 | 0.96 | 0.16 |
| TAES | Sens | 39.46 | 37.80 | 10.97 | 12.37 | 2.04 |
| | Spec | 87.49 | 91.29 | 93.51 | 99.22 | 99.90 |
| | FPs | 28.21 | 18.39 | 11.82 | 1.44 | 0.17 |
| | WGT | -38.57 | -16.58 | -26.08 | 2.46 | 0.82 |

Transfer deep learning is the core of the proposed system and helps our approach in two important ways: spatiotemporal interpretation of EEG signals and neural network convergence. The multiresolution convolutional layers in ResNet18 can effectively encode spatial and temporal relationships. Data augmentation in transfer learning makes the training fast and more importantly, it improves the convergence of a large neural network significantly. While the number of samples in our database is relatively large, we do observe overfitting tendencies on the training dataset. Though cross-validation was used, we plan to explore other pretrained ImageNet networks to assess their impact on overfitting and maintain generality.

## Acknowledgments

This material is based upon work supported by the National Science Foundation under Grant No. IIP-1827565. Any opinions, findings, and conclusions or recommendations expressed in this material are those of the author(s) and do not necessarily reflect the views of the National Science Foundation.

## References

[1] "Epilepsy Statistics," 2021. [Online]. Available: *https://www.epilepsy.com/learn/about-epilepsy-basics/epilepsy-statistics* [Accessed: 01-April-2021].

[2] "Epilepsy Data and Statistics," 2021. [Online]. Available: *https://www.cdc.gov/epilepsy/data/index.html* [Accessed: 01-April-2021].

[3] Y. Roy, R. Iskander, and J. Picone, "The Neureka™ 2020 Epilepsy Challenge," *NeuroTechX*, 2020. [Online]. Available: *https://neureka-challenge.com/*. [Accessed: 16-Apr-2020].

[4] V. Shah, I. Obeid, J. Picone, G. Ekladious, R. Iskander, and Y. Roy, "Validation of Temporal Scoring Metrics for Automatic Seizure Detection," in *Proceedings of the IEEE Signal Processing in Medicine and Biology Symposium (SPMB)*, 2020, pp. 1-5. *https://ieeexplore.ieee.org/abstract/document/9353631*.

[5] A. Sharmila and P. Geethanjali, "A review on the pattern detection methods for epilepsy seizure detection from EEG signals," *Biomed. Eng. / Biomed. Tech.*, vol. 64, no. 5, pp. 507–517, Sep. 2019. *https://doi.org/10.1515/bmt-2017-0233*.

[6] M. Golmohammadi, *Deep Architectures for Spatio-Temporal Sequence Recognition With Applications in Automatic Seizure Detection*, Temple University, 2021. *https://www.isip.piconepress.com/publications/phd_dissertations/2021/seizure_detection/*.

[7] V. Shah, *Improved Segmentation for Automated Seizure Detection Using Channel-Depedent Posteriors*, Temple University, 2021. *https://www.isip.piconepress.com/publications/phd_dissertations/2021/seizure_segmentation/*.

[8] J. Craley, E. Johnson, and A. Venkataraman, "Integrating Convolutional Neural Networks and Probabilistic Graphical Modeling for Epileptic Seizure Detection in Multichannel EEG," in *Information Processing in Medical Imaging*, A. C. S. Chung, J. C. Gee, P. A. Yushkevich, and S. Bao, Eds. Cham: Springer International Publishing, 2019, pp. 291-303. *http://link.springer.com/10.1007/978-3-030-20351-1_22*.

[9] J. Craley, E. Johnson, and A. Venkataraman, "A Novel Method for Epileptic Seizure Detection Using Coupled Hidden Markov Models," in *Lecture Notes in Computer Science (including subseries Lecture Notes in Artificial Intelligence and Lecture Notes in Bioinformatics)*, vol. 11492 LNCS, 2018, pp. 482-489. *http://link.springer.com/10.1007/978-3-030-00931-1_55*.

[10] A. H. Shoeb, *Application of machine learning to epileptic seizure onset detection and treatment*, Harvard University--MIT Division of Health Sciences and Technology, 2009. *https://dspace.mit.edu/handle/1721.1/54669*.

[11] A. Emami, N. Kunii, T. Matsuo, T. Shinozaki, K. Kawai, and H. Takahashi, "Seizure detection by convolutional neural network-based analysis of scalp electroencephalography plot images," *NeuroImage Clin.*, vol. 22, p. 101684, 2019. *https://www.sciencedirect.com/science/article/pii/S2213158219300348*.

[12] C. Gómez, P. Arbeláez, M. Navarrete, C. Alvarado-Rojas, M. Le Van Quyen, and M. Valderrama, "Automatic seizure detection based on imaged-EEG signals through fully convolutional networks," *Sci. Rep.*, vol. 10, no. 1, p. 21833, 2020. *https://doi.org/10.1038/s41598-020-78784-3*.

[13] M. Ihle *et al.*, "EPILEPSIAE – A European epilepsy database," *Comput. Methods Programs Biomed.*, vol. 106, no. 3, pp. 127-138, 2012. *https://www.sciencedirect.com/science/article/pii/S0169260710002221*.

[14] V. Shah *et al.*, "The Temple University Hospital Seizure Detection Corpus," *Front. Neuroinform.*, vol. 12, pp. 1-6, 2018. *http://journal.frontiersin.org/researchtopic/1563/pdf*.

[15] H. A. Haider *et al.*, "Sensitivity of quantitative EEG for seizure identification in the intensive care unit," *Neurology*, vol. 87, no. 9, pp. 935–944, 2016. *http://n.neurology.org/content/87/9/935.long*.

[16] J. Deng, W. Dong, R. Socher, L. Li, K. Li, and L. Fei-Fei, "ImageNet: A large-scale hierarchical image database," in *Proceedings of the IEEE Conference on Computer Vision and Pattern Recognition*, 2009, pp. 248-255. *https://ieeexplore.ieee.org/document/5206848*.

[17] K. He, X. Zhang, S. Ren, and J. Sun, "Deep residual learning for image recognition," in *Proceedings of the IEEE Computer Society Conference on Computer Vision and Pattern Recognition*, 2016, vol. 2016-Decem, pp. 770-778. *http://ieeexplore.ieee.org/document/7780459/*.

[18] The HDF Group, "Learning HDF5," *The HDF Group Documentation*, 2021. [Online]. Available: *https://portal.hdfgroup.org/display/HDF5/Learning+HDF5*. [Accessed: 17-Oct-2021].

[19] S. Lopez, M. Golmohammadi, I. Obeid, and J. Picone, "An analysis of two common reference points for EEGs," in *Proceedings of the IEEE Signal Processing in Medicine and Biology Symposium*, 2016, pp. 1-4. *https://ieeexplore.ieee.org/document/7846854*.

[20] R. C. Gonzalez and R. E. Woods, *Digital Image Processing*, 4th ed. New York City, New York, USA: Pearson, 2017. *https://www.pearson.com/us/higher-education/program/Gonzalez-Digital-Image-Processing-4th-Edition/PGM241219.html*.

[21] V. Shah, M. Golmohammadi, I. Obeid, and J. Picone, "Objective Evaluation Metrics for Automatic Classification of EEG Events," in *Biomedical Signal Processing: Innovation and Applications*, 1st ed., I. Obeid, I. Selesnick, and J. Picone, Eds. New York City, New York, USA: Springer, 2021, pp. 223-256. *https://www.springer.com/gp/book/9783030674939*.